\documentclass[,12pt]{article}
\pdfoutput=1
\usepackage{amsfonts}
\usepackage{amssymb}
\usepackage{amsmath}
\usepackage{cases}
\usepackage{caption}
\usepackage{algpseudocodex}
\usepackage[authoryear]{natbib}
\usepackage[]{graphicx}
\usepackage{tikz}	
\usepackage{pgfplots}
\usepackage{verbatim}
\usepackage{multicol}
\usepackage{cite}
\usepackage[utf8]{inputenc}
\usepackage{mathrsfs}
\usepackage{xfrac}
\usepackage{enumerate}
\usetikzlibrary{arrows}
\usepackage{url}
\usepackage{setspace}
\usepackage{etoolbox}
\usepackage{multirow, makecell}
\pgfplotsset{compat=1.15}
\usepackage{lineno}
\newcommand{\pkg}[1]{{\normalfont\fontseries{b}\selectfont #1}}
\let\proglang=\textsf \let\code=\texttt

\begin{document}

\begin{titlepage}
   \begin{center}
       \vspace*{1cm}

       \textbf{On an EM-based closed-form solution for 2 parameter IRT models}

       \vspace{1.5cm}

Noventa Stefano\\
Methods Center, Universität Tübingen\\
E-Mail: stefano.noventa@uni-tuebingen.de\\
       \vspace{.5cm}
Roberto Faleh\\
Methods Center, Universität Tübingen\\
E-Mail: roberto.faleh@uni-tuebingen.de\\
\vspace{.5cm}
Augustin Kelava\\
Methods Center, Universität Tübingen\\
E-Mail:  augustin.kelava@uni-tuebingen.de\\
\begin{abstract}
It is a well-known issue that in Item Response Theory models there is no closed-form for the maximum likelihood estimators of the item parameters. Parameter estimation is therefore typically achieved by means of numerical methods like gradient search. The present work has a two-fold aim: On the one hand, we revise the fundamental notions associated to the item parameter estimation in 2 parameter Item Response Theory models from the perspective of the complete-data likelihood. On the other hand, we argue that, within an Expectation-Maximization approach, a closed-form for discrimination and difficulty parameters can actually be obtained that simply corresponds to the Ordinary Least Square solution.

\textbf{keywords}: Item Response Theory, EM, parameter estimation, 2 parameter models, OLS

\end{abstract}
            
        \end{center}
\end{titlepage}

\section{Introduction}

Item parameters estimation in Item Response theory (IRT) has a long tradition and a rather extensive literature ranging from Frequentist to Bayesian approaches and spanning different methods and techniques \citep[see, e.g., Section III: Parameter Estimation, in][for an overview]{VII2016}. In their seminal work on the Marginal Maximum Likelihood (MML) approach to item parameters estimation in the normal ogive model, \citet{BL1970} distinguished between unconditional and conditional approaches. The former concerns parameter estimates obtained by integrating over the distribution of the latent ability used to capture a specific population from which a sample of individuals is drawn, while the latter concerns estimates obtained from ``arbitrarily given'' individuals. In a more modern perspective, these corresponds to the treatment of incidental parameters as random vs.\ fixed effects. Interestingly, as an example of unconditional approach \citet{BL1970} indicate the ``heuristic approach'' of \citet{L1944} in which the item easiness is estimated by the normal deviate associated to the percent of correct responses in the sample, while the item discrimination is given by the item loading in a one-dimensional factor analysis based on the matrix of the sample tetrachoric correlations. 

Due to the limited computational feasibility of the MML method, after the seminal contribution of \citet{BL1970}, estimation methods in IRT and Structural Equation Models (SEMs) departed from each other. On the one hand, the unconditional heuristic method further developed in the now well-established approach to measurement models in SEMs in presence of categorical variables, given by the seminal works of \citet{C1975} and \citet{M1978}, which extended the heuristic approach to multiple factors and provided easier computation methods based on the first and second order proportions and involving the use of Generalized Least Square (GLS) estimators. It is indeed now a well-established result that IRT models are equivalent to factor analysis models \citep[see, e.g.,][]{TDL1987}. On the other hand, the MML approach of \citet{BL1970} was further developed by \citet{B1972} and recast by \citet{BA1981} in an easier to apply quadrature-based version of the same procedure. Later refinements were discussed by several authors \citep[see, e.g.,][]{T1982, RT1983, M1992}. Most importantly, these approaches might be re-framed as applications of an Expectation-Maximization \citep[EM,][]{DLR1977, MK1997} algorithm in which, by extending the missing information principle of \citet{OW1972}, both the observed response patterns and the missing latent ability values constitute the complete data.  The procedure is summarized in the left panel of Table \ref{tab:procedures}, for details see Sections \ref{sec:preliminaries} and \ref{sec:completeIRT}. In the expectation step, the posterior distribution of the latent ability conditional to the provisional values of the item parameters is estimated and is used to compute the posterior expectation of the complete-data log-likelihood, which is then optimized in the maximization step. Typically, the latter corresponds to the application of numerical optimization methods like Newton-Raphson to find the solutions of the first partial derivatives of the log-likelihood function. For very complete and general reviews of these procedures the reader is also referred to \citet{HBZ1988} and \citet{WH1996}.

\begin{center}
\captionof{table}{Pseudo-code of the complete-data log-likelihood EM procedure (left) and of the suggested OLS-based EM procedure (right).}
\label{tab:procedures}
\begin{tabular}{cc}
  \begin{minipage}{.5\textwidth}
\scriptsize
\begin{algorithmic}[1]
\State \textbf{Complete-data $(\boldsymbol{X}, \boldsymbol{\theta})$ EM}
\State Initialization and data input ($\boldsymbol{X}$)
\While{Conditions} 
    \State Set counters
    \State Set provisional parameters $\Gamma_0$
    \Procedure{E-step}{}
      \State Estimate Posterior of ability $P(\boldsymbol{\theta}|\boldsymbol{X},\Gamma_0)$
      \LComment{Eq. \eqref{posteriorIRT}}
      \State Estimate Proportions of correct and total responses at a given ability level
       \LComment{Eq.s \eqref{patternbargen} and \eqref{totalbargen}}
      \State Estimate Posterior Expectation of the complete- data log-likelihood $Q = E[\ell_c(\Gamma|\boldsymbol{X},\boldsymbol{\theta})|\boldsymbol{X},\Gamma_0]$
      \LComment{Eq. \eqref{Qone}}
    \EndProcedure
     \Procedure{M-step}{}
     \State Optimize $Q$-function  with gradient search
     \LComment{Eq. \eqref{Qonettraditional}}
    \EndProcedure
    \State Update counters and parameters
    \State Check conditions
 \EndWhile 
 \State Return optimal parameters
\end{algorithmic}
\end{minipage}   & 
\begin{minipage}{0.45\textwidth}\scriptsize
\begin{algorithmic}[1]
\State \textbf{OLS EM}
\State Initialization and data input ($\boldsymbol{X}$)
\While{Conditions} 
    \State Set counters
    \State Set provisional parameters $\Gamma_0$
    \Procedure{E-step}{}
    \State Estimate Posterior of ability $P(\boldsymbol{\theta}|\boldsymbol{X},\Gamma_0)$
     \LComment{Eq. \eqref{posteriorIRT}}
        \State Estimate Proportions of correct and total responses at a given ability level
          \LComment{Eq.s \eqref{patternbargen} and \eqref{totalbargen}}
        \State Estimate Latent Response Variables based on proportions of correct and total responses
         \LComment{Eq. \eqref{OLS}}
    \EndProcedure
     \Procedure{M-step}{}
     \State  OLS estimators for IRT parameters
     \LComment{Eq.s \eqref{OLSestima} and \eqref{OLSestimb}}
    \EndProcedure
    \State Update counters and parameters
    \State Check conditions
 \EndWhile 
 \State Return optimal parameters
\end{algorithmic}
\end{minipage}
\end{tabular}
\end{center}
\vspace{.6cm}

In the present work, by mixing the two traditions discussed above, we argue that actual estimates of the latent response variables underlying the factor analytical approach to categorical indicators can indeed by obtained by means of the EM algorithm. The procedure is summarized in the right panel of Table \ref{tab:procedures}, for details see Section \ref{sec:completeIRT}. As a consequence, discrimination and difficulty parameters can be directly estimated by a simple Ordinary Least Square (OLS) approach, which thus provides a sequence of closed-form estimators to said parameters that converges to the true values. Most of all, such a case corresponds to a `trivial' solution of the first partial derivatives of the log-likelihood function of IRT models in the standard MML approach. This suggests that there might be no need for gradient search methods. Moreover, although limited by the relatively slow performance of the EM, a similar approach might strongly reduce the computational time of the parameter estimation and possibly more easily scale to larger sets of items.

As to the plan of this work, we first introduce in Section \ref{sec:IRT} the 2 parameter  models and the associated observed-data likelihood and log-likelihood, in Section \ref{sec:preliminaries} we recall some notions on the EM procedure and we revise the complete-data likelihood and log-likelihood, in Section \ref{sec:completeIRT} we argue how the system of equations associated to the partial derivatives of the posterior expectation of the complete log-likelihood, which is traditionally solved by means of gradient search methods, can actually be solved by means of a sequence of closed-form OLS estimators. Finally, in Section \ref{sec:simulations}, as a proof of concept, we provide some simulations showing that the estimates of the closed-forms estimators are compatible with those returned by the package \pkg{mirt} \citep{C2012}. A brief exploration of the effects of the number of quadrature points on the routine is given in Appendix \ref{sec:quadsim}. The routine is implemented in the \proglang{R} environment \citep{R2024} and the \proglang{R} script is available upon request.

\section{IRT 2-parameter models}\label{sec:IRT}

Let $X\in\{0,1\}^I$ represent a response pattern in the data collecting the dichotomous responses $X_i\in\{0,1\}$ of $N$ individuals to a set of items $i\in \{1,\ldots, I\}$. The probability of a correct response $X_i=1$ to the $i$-th item, conditional to the ability of the individual $\theta\in\mathbb{R}$ is often given by the Birnbaum model, i.e., a logistic function of the form 
\begin{align}\label{2PL}
P(X_i = 1| \theta, \Gamma_i) & = \frac{e^{a_i(\theta-b_i)}}{1+e^{a_i(\theta-b_i)}} = \frac{e^{a_i\theta+\tau_i}}{1+e^{a_i\theta+\tau_i}}
\end{align}
where $\Gamma_i =\{a_i, b_i\}$ or equivalently $\Gamma_i=\{a_i, \tau_i\}$ contains the discrimination $a_i$, difficulty $b_i$, or threshold $\tau_i$ parameters for the $i$-th item. Although both notations are common in IRT, the first notation is the traditional one of IRT models, while the second one is more typical of factor analysis. Clearly, other functional forms of the 2 parameter model are generally possible as long as they are written in a form like $F(\theta, \Gamma_i)$ for a strictly monotonic increasing function $F: \mathbb{R}\rightarrow [0,1]$, e.g., normal ogive models if $F=\phi$ is the standardized normal cumulative distribution function. By the traditional assumption of local independence (LI) the probability of a given pattern of responses, conditional to the ability of the individual $\theta\in\mathbb{R}$, 
is then given by a conditional probabilistic assumption of the form
\begin{align}\label{local}
P(X|\theta, \Gamma) & = \prod_{i=1}^I P(X_i|\theta, \Gamma_i)  =  \prod_{i=1}^I P(X_i=1|\theta, \Gamma_i)^{X_i}P(X_i=0|\theta, \Gamma_i)^{1-X_i} 
\end{align}
where $\Gamma$ collects all set of parameters $\Gamma_i$ for every item. The observed-data likelihood of a pattern of responses, once marginalized over the latent trait, is then given by 
\begin{align}\label{IRTlikelihood}
L(\Gamma, \mu, \sigma|X) & = P(X|\Gamma, \mu, \sigma) = \int P(X|\theta,\Gamma) f_{\theta}(\theta; \mu, \sigma)d\theta
\end{align}
with ability distribution $f_\theta(\theta;\mu, \sigma)$. In what follows we will assume a standard normal distribution so that $\mu = 0$ and $\sigma = 1$, and these parameters will be omitted from the likelihood expression. Let then $\boldsymbol{X}$ represent the collection of all response patterns $X\in\{0,1\}^I$. The observed-data likelihood and log-likelihood are then generally given (least of some proportional terms like multinomial coefficients) by 
\begin{align}
L(\Gamma|\boldsymbol{X}) & = P(\boldsymbol{X}|\Gamma) = \prod_{X\in\{0,1\}^I}
 P(X|\Gamma)^{N_X} \label{likelihood}\\
\ell(\Gamma|\boldsymbol{X})  & =  \log{L(\Gamma|\boldsymbol{X})} = \log{P(\boldsymbol{X}|\Gamma)}\nonumber\\
& = \sum_{X\in\{0,1\}^I}N_X\log{P(X|\Gamma)} = \sum_{X\in\{0,1\}^I}N_X \ell(\Gamma|X) \label{loglikelihood}
\end{align}
where $N_X$ is the frequency of persons in pattern $X$. If a random sampling view \citep{H1990} is assumed, the likelihood \eqref{IRTlikelihood} expresses the proportion of all individuals $P(X|\Gamma)$ in a population that would provide response $X$, so the multinomial-like observed-data likelihood \eqref{likelihood} expresses the probability of drawing $N_X$ individuals with probability $P(X|\Gamma)$ for all $X\in\{0,1\}^I$. In what follows, we will need to consider discrete latent ability classes for mostly two purposes: on the one hand, they are used in numerical approximation methods (e.g., Gauss-Hermite); on the other hand, they will be an essential part of the construction of EM-based closed-forms estimators. If we then consider $T$ discrete latent ability classes, we can write the observed-data likelihood and log-likelihood as
\begin{align}
L(\Gamma|\boldsymbol{X}) & = \prod_{X\in \{0,1\}^{I}}( \sum_{t=1}^T P(X|\theta_t,\Gamma)\nu_t)^{N_X}\label{IRTdiscrete}\\
\ell(\Gamma|\boldsymbol{X}) & = \sum_{X\in \{0,1\}^{I}}N_X\log{(\sum_{t=1}^T P(X|\theta_t,\Gamma)\nu_t)}\label{IRTlogdiscrete}
\end{align}
where $\nu_t = \nu(\theta_t)$ is the membership/point mass probability of the latent class value $\theta_t$. In the most general case, the likelihood can be written as $L(\Gamma, \nu|\boldsymbol{X})$ where $\nu$ is the vector of such class membership parameters. However, typically in IRT these probabilities are not considered parameters to be estimated as they are replaced by quadrature weights $A(\theta_t)$ of the quadrature point $\theta_t$ and can therefore be omitted. It is worth mentioning that, in terms of implementation, the approach introduced by \citet{BA1981} directly searches for the roots of the partial derivatives of the observed-data log-likelihood \eqref{IRTlogdiscrete}. For the interested reader, such an approach is briefly discussed in Appendix \ref{sec:marginalization}. In the present work, however, we follow an equivalent but more general complete-data log-likelihood treatment as discussed by \citet{WH1996}. In order to do so, we need to first briefly review the EM procedure.

\section{A brief recap of the EM algorithm}\label{sec:preliminaries}

In the EM approach it is convenient to consider the complete-data likelihood, for which one hypothesizes that, alongside the responses $\boldsymbol{X}$, the ability parameter $\theta$ is actually measured. In order to distinguish them from the observed-data likelihood $L$ and log-likelihood $\ell$ as given by Equations \eqref{IRTdiscrete} and \eqref{IRTlogdiscrete}, we label the complete-data likelihood $L_c$ and log-likelihood $\ell_c$. Given then $T$ ability classes, these  respectively take form
\begin{align}
L_c(\Gamma|\boldsymbol{X},\boldsymbol{\theta}) & = P(\boldsymbol{X},\boldsymbol{\theta}|\Gamma) = \prod_{X\in\{0,1\}^I}\prod_{t=1}^TP(X, \theta_t|\Gamma)^{N_{Xt}}\label{IRTcompletelikelihood}\\
\ell_c(\Gamma|\boldsymbol{X},\boldsymbol{\theta})  & = \log L_c(\Gamma|\boldsymbol{X},\boldsymbol{\theta}) =\sum_{X\in\{0,1\}^I} \sum_{t=1}^T N_{Xt} \log P(X, \theta_t|\Gamma)\label{IRTcompleteloglikelihood}
\end{align}
where the unknown quantity $N_{Xt}$ captures the number of individuals within ability class $t$ and with response pattern $X$, and where $\boldsymbol{\theta}$ is the vector of all ability values $\theta_t$ for $t\in \{1,\ldots,T\}$. Clearly, the complete-data likelihoods \eqref{IRTcompletelikelihood} an \eqref{IRTcompleteloglikelihood} differ from the observed-data likelihoods \eqref{likelihood}
and \eqref{loglikelihood}. The basic idea of the EM procedure is that an iterative approach to estimate the parameters $\Gamma$ via the complete-data log-likelihood will return the same estimates of the MML solution for the observed-data log-likelihood. Since by definition it holds that $P(X,\theta_t |\Gamma)=P(\theta_t|X,\Gamma)P(X|\Gamma)$ we can write
\begin{align}\label{basic}
\ell(\Gamma|X) = \log{P(X|\Gamma)} &= \log{P(X,\theta_t |\Gamma)}- \log{P(\theta_t|X,\Gamma)}\nonumber\\
& = \ell_c(\Gamma|X, \theta_t)- \log{P(\theta_t|X,\Gamma)}.
\end{align}

Since $\sum_{t=1}^T N_{Xt} = N_X$ and thus $\sum_{X,t}N_{Xt}\ell(\Gamma|X) =\sum_{X}(\sum_{t}N_{Xt})\ell(\Gamma|X) = \ell(\Gamma|\boldsymbol{X})$ by Equation \eqref{loglikelihood}, multiplication on both sides of Equation \eqref{basic} by $N_{Xt}$ and summation over $X\in \{0,1\}^{I}$ and $t\in\{1,\ldots, T\}$ yields the following relation between observed-data and complete-data log-likelihoods
\begin{align}\label{likelihoodrel}
\ell(\Gamma|\boldsymbol{X}) &= \ell_c(\Gamma|\boldsymbol{X},\boldsymbol{\theta}) - \sum_{X\in \{0,1\}^{I}} \sum_{t=1}^T N_{Xt}\log{P(\theta_t|X,\Gamma)}\nonumber\\
& = \ell_c(\Gamma|\boldsymbol{X},\boldsymbol{\theta})-\log{P(\boldsymbol{\theta}|\boldsymbol{X},\Gamma)}.
\end{align}

At the same time, if we take the posterior expectation $E[\boldsymbol{\cdot}|X,\Gamma_0]$ of \eqref{basic} over the missing data $\theta$, conditional to $X$ and a given provisional set of parameters $\Gamma_0$, we obtain
\begin{align*}
\ell(\Gamma|X) & = E[\log{P(X, \theta_t|\Gamma)}-\log{P(\theta_t|X,\Gamma)}|X,\Gamma_0]\\
 & = \sum_{t = 1}^T P(\theta_t|X,\Gamma_0)[\log{P(X, \theta_t|\Gamma)} - \log{P(\theta_t|X,\Gamma)}]
\end{align*}
since $\sum_{t=1}^TP(\theta_t|X, \Gamma_0)\ell(\Gamma|X)=\ell(\Gamma|X)$. Further multiplication by $N_X$ and summation over $X\in \{0,1\}^{I}$ yields the following relation
\begin{align}\label{likelihoodrel2}
\ell(\Gamma|\boldsymbol{X}) & = \sum_{X\in \{0,1\}^{I}}\sum_{t=1}^TN_XP(\theta_t|X,\Gamma_0)[\log{P(X, \theta_t|\Gamma)}-\log{P(\theta_t|X,\Gamma)}]
\end{align}
which by comparison with relation \eqref{likelihoodrel} allows the identification
\begin{align}\label{QEM}
N_{Xt} & = N_XP(\theta_t|X,\Gamma_0),
\end{align}
which provides an estimate of $N_{Xt}$ based on the provisional parameters $\Gamma_0$ of the number of individuals with ability level $\theta_t$ that provide pattern of responses $X$. In order to better understand now the EM iterative process, let us first rewrite relation \eqref{likelihoodrel2} as
\begin{align}\label{posterior expectation}
\ell(\Gamma|\boldsymbol{X}) & = E[\ell_c(\Gamma|\boldsymbol{X},\boldsymbol{\theta})|\boldsymbol{X},\Gamma_0]-E[\log{P(\boldsymbol{\theta}|\boldsymbol{X},\Gamma))}|\boldsymbol{X},\Gamma_0]\nonumber\\
&  = Q(\Gamma|\Gamma_0)+H(\Gamma|\Gamma_0)
\end{align}
where the posterior expectation $Q(\Gamma|\Gamma_0)$ is calculated in the E-step and maximized in the M-step of an EM. As $H(\Gamma|\Gamma_0)$ is the cross-entropy of the posterior distributions $P(\boldsymbol{\theta}|\boldsymbol{X},\Gamma)$ and $P(\boldsymbol{\theta}|\boldsymbol{X},\Gamma_0)$ it must hold (Gibb's inequality) that $H(\Gamma_0|\Gamma_0)\leq H(\Gamma|\Gamma_0)$ for all $\Gamma$ since $H(\Gamma_0|\Gamma_0)$ is the entropy of $P(\boldsymbol{\theta}|\boldsymbol{X},\Gamma_0)$. As a consequence, it must also hold that $\ell(\Gamma|\boldsymbol{X})-\ell(\Gamma_0|\boldsymbol{X}) \geq Q(\Gamma|\Gamma_0)- Q(\Gamma_0|\Gamma_0)$
so that any increase in $Q$ (i.e, the posterior expectation of the complete log-likelihood $\ell_c(\Gamma|\boldsymbol{X},\boldsymbol{\theta})$) is a lower bound for the increase in the observed-data log-likelihood $\ell(\Gamma|\boldsymbol{X})$. In the EM method, the $n$-th E-step is the calculation of $P(\boldsymbol{\theta}|\boldsymbol{X},\Gamma^{(n)})$, of all of the $N_{Xt}^{(n)}=N_XP(\theta_t|X, \Gamma^{(n)})$, and of $Q(\Gamma|\Gamma^{(n)})$ for a provisional set of parameters $\Gamma^{(n)}$, while the M-step is the maximization of $Q(\Gamma|\Gamma^{(n)})$ w.r.t.\ the parameters $\Gamma$ to find a new set $\Gamma^{(n+1)}$ for the next iteration. The maximization step is sometimes substituted with finding the roots of the partial derivatives if these have a closed-form. However, the EM algorithm does not always guarantee convergence to the global maximum of the likelihood. For this reason, it is common practice to run the algorithm multiple times from different initial values. For more details concerning the convergence of the EM see, e.g., \citet{W1983}.

\section{Complete-data likelihood for discrete IRT models}\label{sec:completeIRT}

In the present section, we first review the complete-data likelihood derivation of the system of equations that yields the ML solution for an IRT model. Subsequently, in place of the traditional solution based on gradient search methods, a new analytical perspective is suggested that yields a sequence of EM-based closed form estimators. 

\subsection{Complete-data likelihood and the traditional solution to ML estimates}

As discussed by \citet{WH1996}, the $Q$-function \eqref{posterior expectation} allows to estimate all parameters $\Gamma$ and $\nu$. Indeed, since $P(X,\theta_t|\Gamma) = P(X|\theta_t,\Gamma)\nu_t$, the posterior expectation of the complete-data log-likelihood conditional to the observed responses and given the estimate of the parameters at the previous $n$-th step, can be split into two terms
\begin{align}
Q(\Gamma|\Gamma^{(n)}) & = E[\ell_c(\Gamma|\boldsymbol{X},\boldsymbol{\theta})|\boldsymbol{X},\Gamma^{(n)}] = E[\log(P(\boldsymbol{X},\boldsymbol{\theta}|\Gamma))| \boldsymbol{X}, \Gamma^{(n)}]\nonumber\\
  & = \sum_{X\in\{0,1\}^I}\sum_{t=1}^TN_XP(\theta_t|X,\Gamma^{(n)}) \log(P(X|\theta_t, \Gamma)\nu_t)\nonumber\\
&  = Q_1(\Gamma|\Gamma^{(n)})+Q_2(\nu|\Gamma^{(n)}),
\end{align}
which are respectively given by 
\begin{align}
Q_1(\Gamma|\Gamma^{(n)}) & = \sum_{X\in\{0,1\}^I}\sum_{t=1}^TN_XP(\theta_t|X,\Gamma^{(n)}) \log{P(X|\theta_t, \Gamma)}\label{finitemixtureQ1}\\
Q_2(\nu|\Gamma^{(n)}) & = \sum_{X\in\{0,1\}^I}\sum_{t=1}^TN_XP(\theta_t|X,\Gamma^{(n)})\log{\nu_t}\label{finitemixtureQ2}
\end{align}
and that separate the dependence of the $Q$-function on the item parameters $\Gamma$ and on the membership probabilities $\nu$. Specifically, the first term $Q_1(\Gamma|\Gamma^{(n)})$ simplifies when taking the partial derivatives w.r.t.\ to the probabilities $\nu$ since it does not depend on $\nu$ but only on the parameters in $\Gamma$, while the second term $Q_2(\nu|\Gamma^{(n)})$ simplifies when taking the partial derivatives w.r.t.\ to the item parameters $\Gamma$ since it does not depend on said parameters but only on $\nu$. In what follows, however, we are only interested in the item parameters $\Gamma$, since the membership parameters $\nu$ are replaced by quadrature weights $A(\theta_t)$ and are therefore considered constants (for a derivation of these parameters see Appendix \ref{sec:memnership}). As a consequence, we only consider Equation \eqref{finitemixtureQ1} and we can also rewrite the posterior probability $P(\theta_t| X, \Gamma^{(n)})$ as 
\begin{align}\label{posteriorIRT}
    P(\theta_t| X, \Gamma^{(n)}) & = \frac{P(X, \theta_t|\Gamma^{(n)})}{P(X| \Gamma^{(n)})} = \frac{P(X|\theta_t,\Gamma^{(n)})\nu_{t}}{\sum\limits_{t'=1}^TP(X|\theta_{t'},\Gamma^{(n)})\nu_{t'}} \nonumber\\
& = \frac{P(X|\theta_t,\Gamma^{(n)})A(\theta_t)}{\sum\limits_{t'=1}^TP(X|\theta_{t'},\Gamma^{(n)})A(\theta_{t'})}
\end{align}
where the last passage is the numerical quadrature case. Clearly, if the membership probabilities are not treated as constants then they must be replaced by the values $\nu^{(n)}_t$ at the $n$-th iteration. Let $P_{i}(\theta_t, \Gamma_i)=P(X_i=1|\theta_t, \Gamma_i)$ and $Q_{i}(\theta_t, \Gamma_i)=1-P(X_i=1|\theta_t, \Gamma_i)$.  By considering now Equation \eqref{finitemixtureQ1} and plugging in local independence \eqref{local} one obtains the following posterior expectation of the complete-data log-likelihood
{\small\begin{align}
   Q_1(\Gamma|\Gamma^{(n)}) & = \sum_{X\in\{0,1\}^I} \sum_{t=1}^T N_X P(\theta_t| X, \Gamma^{(n)}) \log{(
   \prod_{i=1}^I P_{i}(\theta_t, \Gamma_i)^{X_{i}}Q_{i}(\theta_t, \Gamma_i)^{(1-X_{i})})}\nonumber \\
    & = \sum_{t=1}^T \sum_{X\in\{0,1\}^I} N_X P(\theta_t| X, \Gamma^{(n)})[\sum_{i=1}^I(X_i\log{P_{i}(\theta_t, \Gamma_i)}\nonumber\\
    & +(1-X_i)\log{Q_{i}(\theta_t, \Gamma_i)})]\nonumber \\
   & = \sum_{t=1}^T \sum_{i=1}^I 
   [N^{1,(n)}_{it}\log(P_{i}(\theta_t, \Gamma_i)) +  (N^{(n)}_{t}-N^{1,(n)}_{it})\log{Q_{i}(\theta_t, \Gamma_i)}]\label{Qone}
\end{align}}
where by using the posterior \eqref{posteriorIRT} in the last passage we have set 
{\small\begin{align}
   N^{1,(n)}_{it}  & = \sum_{X\in\{0,1\}^I}X_iN_XP(\theta_t| X, \Gamma^{(n)}) = \sum_{X\in\{0,1\}^I}
\frac{X_iN_XP(X|\theta_t,\Gamma^{(n)})\nu_t}{\sum_{t'}P(X|\theta_{t'},\Gamma^{(n)})\nu_{t'}}\label{patternbargen}\\
   N^{(n)}_{t}   & = \sum_{X\in\{0,1\}^I}N_XP(\theta_t| X, \Gamma^{(n)}) = \sum_{X\in\{0,1\}^I} \frac{N_X P(X|\theta_t,\Gamma^{(n)})\nu_t}{\sum_{t'}P(X|\theta_{t'},\Gamma^{(n)})\nu_{t'}},\label{totalbargen}
\end{align}}
which can be interpreted in the following way. $N^{(n)}_t$ is the proportion of the total number $N$ of individuals in the quadrature point $\theta_t$ at the $n$-th iteration, so that $\sum_{t=1}^T N^{(n)}_t=N$ at every step. Similarly, $N^{1,(n)}_{it}$ can be interpreted as the proportion of individuals at the quadrature point $\theta_t$ and at the $n$-th step that provide a correct response to the $i$-th item. One can then define the proportion $N^{0,(n)}_{it}=N_{t}-N^{1,(n)}_{it}$ of individuals at the quadrature point $\theta_t$ and at the $n$-th step that provide an incorrect response to the $i$-th item, which corresponds to replace  $X_i$ with $(1-X_i)$. By taking the partial derivative of \eqref{Qone} w.r.t.\ an arbitrary parameter $\gamma_j\in\Gamma_j$, one obtains the following general expression
{\small\begin{align}\label{Qonepartial}
\frac{\partial Q_1(\Gamma|\Gamma^{(n)}) }{\partial\gamma_j} & = \sum_{t=1}^T \left(\frac{N^{1,(n)}_{jt}}{P_{j}(\theta_t, \Gamma_j)}-\frac{N^{0,(n)}_{jt}}{Q_{j}(\theta_t, \Gamma_j)}\right)\frac{\partial P_{j}(\theta_t, \Gamma_j)}{\partial\gamma_j} = \sum_{t=1}^T \phi^{(n)}_{jt} \frac{\partial P_{j}(\theta_t, \Gamma_j)}{\partial\gamma_j},
\end{align}}
which is a system of equations (one for every $\gamma_j\in\Gamma$ and for every item) that once set to zero provides the MML solution. Notice that we have set $\phi^{(n)}_{jt}= \frac{N^{1,(n)}_{jt}}{P_{j}(\theta_t, \Gamma_j)}-\frac{N^{0,(n)}_{jt}}{Q_{j}(\theta_t, \Gamma_j)}$. Traditionally, as in \citet{BA1981}, such a system of equations is rewritten as
\begin{align}\label{Qonettraditional}
\frac{\partial Q_1(\Gamma|\Gamma^{(n)}) }{\partial\gamma_j} & = \sum_{t=1}^T \left(\frac{N^{1,(n)}_{jt}-N^{(n)}_{t}P_{j}(\theta_t, \Gamma_j)}{P_{j}(\theta_t, \Gamma_j)Q_{j}(\theta_t, \Gamma_j)}\right)\frac{\partial P_{j}(\theta_t, \Gamma_j)}{\partial\gamma_j} 
\end{align}
and is solved by means of gradient search methods (see also Appendix \ref{sec:marginalization} for further details on the derivation of Equation \eqref{Qonettraditional} from the observed-data log-likelihood). In what follows, we argue that gradient search methods might in principle not be needed to solve the previous system of equations and we suggest a new analytical perspective to its solution.

\subsection{EM-based closed form estimators for IRT parameters}

In the present subsection we suggest a new approach to the solution of the system of equations \eqref{Qonepartial} by arguing that its solution is meaningful for IRT models (i.e., for the true values $\Gamma$ of the parameters) only in the trivial case in which every $\phi$-term in Equation \eqref{Qonepartial} is identically null, independently on the choice of IRF $P_{j}(\theta_t, \Gamma_j)$ or parameter $\gamma_j\in\Gamma_j$. In other words, system \eqref{Qonepartial} actually contains an equation for every item and every quadrature point, but the equations for every parameter $\gamma_j\in\Gamma_j$ are redundant and ignorable. As a consequence, the $\phi$-terms form a sequence that converges to zero and that can be replaced, via a plug-in approach, by a new sequence of $\phi$-terms that allows to define OLS estimators for the parameters at every step of the EM procedure.

As a first step, let us observe that by following the interpretation of the proportions \eqref{patternbargen} and \eqref{totalbargen} given above, the $n$-th ratio $P^{(n)}_{j}(\theta_t) = N^{1,(n)}_{jt}/N^{(n)}_{t}$ can be interpreted as an approximation at the $n$-th step of the marginal IRF $P_{j}(\theta_t, \Gamma^{(n)}_j)$, with $\Gamma^{(n)}_j=\{a^{(n)}_j, b^{(n)}_j\}$, and converges to the actual IRF $P_{j}(\theta_t, \Gamma_j)$ in the limit $\Gamma^{(n)}\rightarrow \Gamma$ under the assumption of LI. Indeed, since the proportion $N_X$ of individuals in a given pattern $X$ can be assumed to be given by $P(X|\Gamma)N$, with $\Gamma$ the true parameters, then one can rewrite the ratio as
\begin{align}\label{heuristicP}
    P^{(n)}_{j}(\theta_t) & = \frac{N^{1,(n)}_{jt}}{N^{(n)}_{t}}  = \sum_{X\in\{0,1\}^I}\frac{X_jP(\theta_t|X,\Gamma^{(n)})P(X|\Gamma)N}{\sum\limits_{X'\in \{0,1\}^{I}}P(\theta_t|X',\Gamma^{(n)})P(X'|\Gamma)N}
  \end{align}  
so that, once $N$ is simplified and since by definition $\sum_{X\in \{0,1\}^{I}}P(X,\theta_t|\Gamma) = P(\theta_t|\Gamma)$, in the limit in which $\Gamma^{(n)}$ converges to $\Gamma$ one has    
\begin{align}
 \lim_{\Gamma^{(n)}\rightarrow \Gamma}  P^{(n)}_{j}(\theta_t)  & = \sum_{X\in\{0,1\}^I}\frac{X_jP(X,\theta_t|\Gamma)}{\sum\limits_{X'\in \{0,1\}^{I}}P(X',\theta_t|\Gamma)}\nonumber\\
&  = \sum_{X\in\{0,1\}^I}X_jP(X|\theta_t, \Gamma) = P_j(\theta_t, \Gamma_j)\nonumber
  \end{align} 
where the last passage follows from the LI assumption. Besides, that  $P^{(n)}_{j}(\theta_t)$ only approximates $P_{j}(\theta_t, \Gamma^{(n)}_j)$ is evident from the fact that Equation \eqref{heuristicP} contains $\Gamma$ in place of $\Gamma^{(n)}$. Hence, if one were to take the limit $\Gamma\rightarrow\Gamma^{(n)}$ then $P^{(n)}_{j}(\theta_t)$ would converge to $P_{j}(\theta_t, \Gamma^{(n)}_j)$ for the same rationale as above. In other words, $P^{(n)}_{j}(\theta_t)$ acts like a bridge between $P_{j}(\theta_t, \Gamma^{(n)}_j)$ and $P_{j}(\theta_t, \Gamma_j)$. Most importantly, Equation \eqref{Qonepartial} is satisfied for the IRT model (i.e., for the true parameters $\Gamma$) by the trivial solution in which every $\phi$-term is null rather than their sum being null. Indeed, such a case corresponds to the situation in which the $n$-th estimates of the odds converge to the actual odds
\begin{align}\label{odds}
   \lim_{\Gamma^{(n)}\rightarrow \Gamma} \frac{N^{1,(n)}_{jt}}{N^{0,(n)}_{jt}} = \lim_{\Gamma^{(n)}\rightarrow \Gamma} \frac{\frac{N^{1,(n)}_{jt}}{N^{(n)}_{t}}}{1-\frac{N^{1,(n)}_{jt}}{N^{(n)}_{t}}} =  \frac{P_{j}(\theta_t, \Gamma_j)}{Q_{j}(\theta_t, \Gamma_j)}\quad\Rightarrow\quad \lim_{\Gamma^{(n)}\rightarrow \Gamma} \phi^{(n)}_{jt} = 0,
\end{align}
so that the sequence of $\phi^{(n)}_{jt}$ terms converges to zero. As the MML solution occurs at the trivial case in which all $\phi$-terms are null, at the actual solution point there is in essence no trade-offs between estimates of the odds and the actual odds at the different values of abilities, which instead occurs at any other $n$-th step of the EM. This can be easily seen with two quadrature points, for which one has that $\phi^{(n)}_{j1}\frac{\partial P_{j}(\theta_1, \Gamma_j)}{\partial\gamma_j} = - \phi^{(n)}_{j2}\frac{\partial P_{j}(\theta_2, \Gamma_j)}{\partial\gamma_j}$. 

A relevant consequence is that Equation \eqref{heuristicP} and the $\phi$-terms can be used to write down closed-form solutions for the discrimination and difficulty parameters in 2 parameter models by plugging-in $P_j(\theta_t, \Gamma^{(n)}_j)$ in place of $P_j(\theta_t, \Gamma_j)$. Indeed, if at every step of the EM one were to consider the IRF $P_j(\theta_t, \Gamma^{(n)}_j)$ and require such an IRF to be given exactly by Equation \eqref{heuristicP}, that is an ideal case in which $P_j(\theta_t, \Gamma^{(n)}_j)=P^{(n)}_{j}(\theta_t)$, this would yield the following system of equations
\begin{align}\label{newequations}
    N^{1,(n)}_{jt}-N^{(n)}_t P_j(\theta_t, \Gamma^{(n)}_j) & = 0.
\end{align}
This requirement would be equivalent to plug-in $\Gamma^{(n)}_j$ within both $P_{j}(\theta_t, \Gamma_j)$ and $Q_{j}(\theta_t, \Gamma_j)$ and require the associated value of the $\phi$-term to be null at every step of the EM, i.e., to substitute the sequence of $\phi^{(n)}_{jt}$ with the sequence of identically null $\phi$-terms
\begin{align}\label{newphi}
    \Tilde{\phi}^{(n)}_{jt} & = \frac{N^{1,(n)}_{jt}}{P_{j}(\theta_t, \Gamma^{(n)}_j)}-\frac{N^{0,(n)}_{jt}}{Q_{j}(\theta_t, \Gamma^{(n)}_j)} = 0,
\end{align}
which would be essentially like forcing the system to obey the IRT condition solution at every step of the EM procedure. A little algebra shows indeed that system \eqref{newphi} and \eqref{newequations} are equivalent. Most of all, as the IRF is of the form $P_j(\theta_t, \Gamma^{(n)}_j)= F(\theta_t, \Gamma^{(n)}_j)$ with $F$ strictly monotonic, it could be inverted to provide at every step of the EM procedure a latent response variable $y^{(n)}_{jt} = F^{-1}(P^{(n)}_j(\theta_t)) = F^{-1}(\frac{N^{1,(n)}_{jt}}{N^{(n)}_{t}})$ for which one could write the traditional regression-like form of factor analysis models. As an example, in the specific case of a 2 parameter logistic model with IRF of the form $F(a^{(n)}_j\theta_t+\tau^{(n)}_j)$ as in Equation \eqref{2PL}, one could then consider the system of equations (one per each quadrature point and each item) given by the linear regressions of the quadrature points on a latent response variable $y_{jt}$ defined by log-odds of the proportions of individuals, that is 
\begin{align}\label{regressionFA}
    y^{(n)}_{jt} = \log{\frac{N^{1,(n)}_{jt}}{N^{(n)}_t-N^{1,(n)}_{jt}}} & =  a^{(n)}_j\theta_t+\tau^{(n)}_j \quad \text{for}\quad t=1,\ldots, T
\end{align}
with discrimination $a^{(n)}_j$ and threshold $\tau^{(n)}_j$ and where the value of the latent response variable $y^{(n)}_{jt}$ is provided by the EM procedure. Since however $P^{(n)}_{j}(\theta_t)$ only approximates $P_j(\theta_t, \Gamma^{(n)}_j)$, one needs to introduce some variability $\epsilon^{(n)}_{jt}$ in Equation \eqref{regressionFA} thus obtaining
\begin{align}\label{OLS}
    y^{(n)}_{jt} = \log{\frac{N^{1,(n)}_{jt}}{N^{(n)}_t-N^{1,(n)}_{jt}}} & =  a^{(n)}_j\theta_t+\tau^{(n)}_j +\epsilon^{(n)}_{jt}\quad \text{for}\quad t=1,\ldots, T
\end{align}
under which the $\phi$-terms \eqref{newphi} can be rewritten as a function of $y^{(n)}_{jt}$ and $\epsilon^{(n)}_{jt}$, that is
{\small\begin{align}
    \Tilde{\phi}^{(n)}_{jt} & = N_t\left[\frac{e^{y^{(n)}_{jt}}(1+e^{a^{(n)}_j\theta_t+\tau^{(n)}_j})}{e^{a^{(n)}_j\theta_t+\tau^{(n)}_j}(1+e^{y^{(n)}_{jt}})}-\frac{1+e^{a^{(n)}_j\theta_t+\tau^{(n)}_j}}{1+e^{y^{(n)}_{jt}}}\right] = N_t\frac{1+e^{y^{(n)}_{jt}-\epsilon^{(n)}_{jt}}}{1+e^{y^{(n)}_{jt}}}\left[e^{\epsilon^{(n)}_{jt}}-1\right]
\end{align}}
and can in principle differ from zero, but must still converge to zero for the true parameters. As the value of the latent response variable $y^{(n)}_{jt}$ is provided by the EM procedure at every step one can then directly consider the OLS solution to Equation \eqref{OLS}, that is
\begin{align}
    \hat{a}^{(n)}_j & = \frac{\sum\limits_{t=1}^T (\theta_t-\overline{\theta})( y^{(n)}_{jt}-\overline{y}^{(n)}_j)}{\sum\limits_{t=1}^T (\theta_t-\overline{\theta})^2}\label{OLSestima}\\
    \hat{\tau}^{(n)}_j & = \overline{y}^{(n)}_j-\hat{a}^{(n)}_j\overline{\theta}\quad \Rightarrow \quad \hat{b}^{(n)}_j = -\frac{\hat{\tau}^{(n)}_j}{\hat{a}^{(n)}_j}\label{OLSestimb}
\end{align}
where $\overline{y}^{(n)}_j$ and $\overline{\theta}$ are the average values of the quadrature points and of the log-odds estimated by means of the posterior in the E-step of the EM. By construction, the sequence of estimators $\hat{\Gamma}^{(n)}_j=\{\hat{a}^{(n)}_j,\hat{\tau}^{(n)}_j,\hat{b}^{(n)}_j\}$ converges to the true values $\Gamma_j = \{a_j,\tau_j, b_j\}$. The effect of the number of quadrature points on the estimates is briefly explored in Appendix \ref{sec:quadsim}. As a final remark, it is important to stress that the sequence of estimators $\hat{\Gamma}^{(n)}$ converges to the true parameters $\Gamma$ as long as $\Gamma$ do actually contains the true parameters. As it can be noticed by Equation \eqref{heuristicP}, what drives the converge of $P^{(n)}(\theta_t)$ to $P(\theta_t, \Gamma_j)$ in the limit $\Gamma^{(n)}\rightarrow \Gamma$ is the implicit assumption that the observed frequencies $N_X$ can be replaced by the proportion $P(X|\Gamma)N$ with $\Gamma$ the true parameters. However, since the observed frequencies $N_X$ are actually expected to be close to the expected frequencies but not to be the same due to noise in the data, in principle $\hat{\Gamma}^{(n)}$ might converge to an alternative set of parameters $\Gamma'$ that actually generates the observed frequencies. Such an issue would not be an issue of local or global identifiability of the model (i.e., structural identifiability), but rather of practical identifiability \citep[see, e.g.,][]{WEA2021} in that different sets of parameters might generate slightly different sets of observed frequencies that are indistinguishable from those generated by the true parameters. This would become even more relevant for more complex models like the 3PL or 4PL in which, as already \citet{M1986} pointed out, the instability of the ML estimates in the 3PL model stems from the fact that different triples $\Gamma_j$ ``can trace similar IRFs in the region of the ability scale where the sample of examinees is to be found'' often resulting in nearly flat likelihood surfaces.

\section{Implementation and simulations}\label{sec:simulations}

For a proof-of-concept implementation of the EM-based OLS approach described in Section \ref{sec:completeIRT} we followed the same Gauss-Hermite quadratures approach applied by \citet{BA1981}, which is routinely used in IRT applications, to replace the membership probabilities $\nu_t$ with some weights $A(\theta_t, \mu, \sigma)$ associated to the quadrature points $\theta_t$. Since for simplicity we consider a standardized normal distribution $f_\theta$ we directly replace $\nu_t$ with the expression $A(\theta_t) = \pi^{-.5}w(\theta_t)$ where $w(\theta_t)$ is the actual quadrature weight at $\theta_t$. Exactly like one needs to additionally rescale the weight by a factor $\pi^{-.5}$, the quadrature points needs also to be rescaled by a factor $\sqrt{2}$ to adjust for the change of variables \citep[see, e.g.,][]{LP1994}. Parameter estimation was carried out for the 1PL and 2PL models with a choice of respectively 2 and 4 quadrature points as these values appear to provide more stable estimates. A brief simulation detailing the effects of the increase in the number of quadrature points on the parameter estimation is given in Appendix \ref{sec:quadsim}.

The \code{EM\_OLS} procedure was implemented in the \proglang{R} environment \citep{R2024}, version 4.2.3. The script for the routine is available upon request to the authors. Gauss-Hermite quadrature points and weights were generated with the package \pkg{fastGHQuad} \citep{B2022}. Resulting parameters estimates of the \code{EM\_OLS} procedure were compared with those returned by \pkg{mirt} \citep{C2012}. Simulated data were generated via the package \pkg{psychotools} \citep{SEA2022}. Simulations were run on a laptop with Intel core i9-12900H processor (2.50 GHz) and 64GB RAM. Parallelization on 16 cores was performed with the packages \pkg{doParallel} \citep{MW2022a}, \pkg{foreach} \citep{MW2022b}, and \pkg{tidyr} \citep{WVG2024}. General run-time was about an hour. For the 1PL, one run of the \code{EM\_OLS} estimation procedure was on average 22.6ms, with min-max range $(16.3,98.62)$, while one run of the \code{mirt} estimation procedure was on average 261.3ms, with min-max range $(39.0,618.0)$. For the 2PL, one run of the \code{EM\_OLS} estimation procedure was on average 52.4ms, with min-max range $(20.3,401.1)$, while one run of the \code{mirt} estimation procedure was on average 154.5ms, with min-max range $(76.9,556.3)$. Benchmarking was performed on 1000 repetitions using the package \pkg{microbenchmark} \citep{M2023}. These comparisons are purely indicative since the \code{EM\_OLS} code is not optimized, while the \code{mirt} routine performs several tasks along with the estimation.

\begin{figure}[h!]
\centering
\includegraphics[scale=.10]{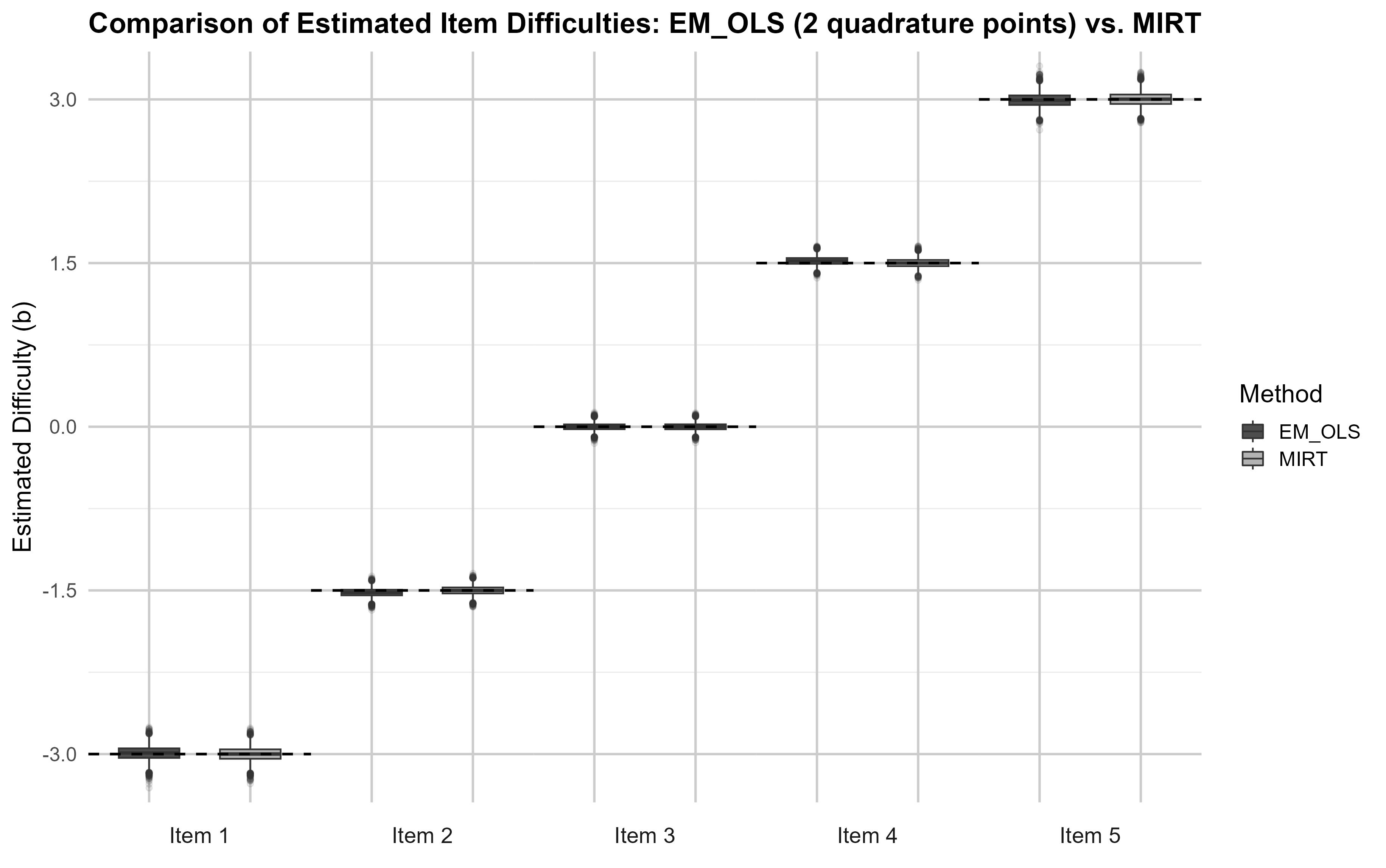}
\caption{Box plots of the estimates of the difficulty parameters for the \code{EM\_OLS}  and \code{mirt} procedures in the 1PL model}
\label{fig:OLSmirt1PL}
\end{figure}

\begin{figure}[h!]
\centering
\includegraphics[scale=.10]{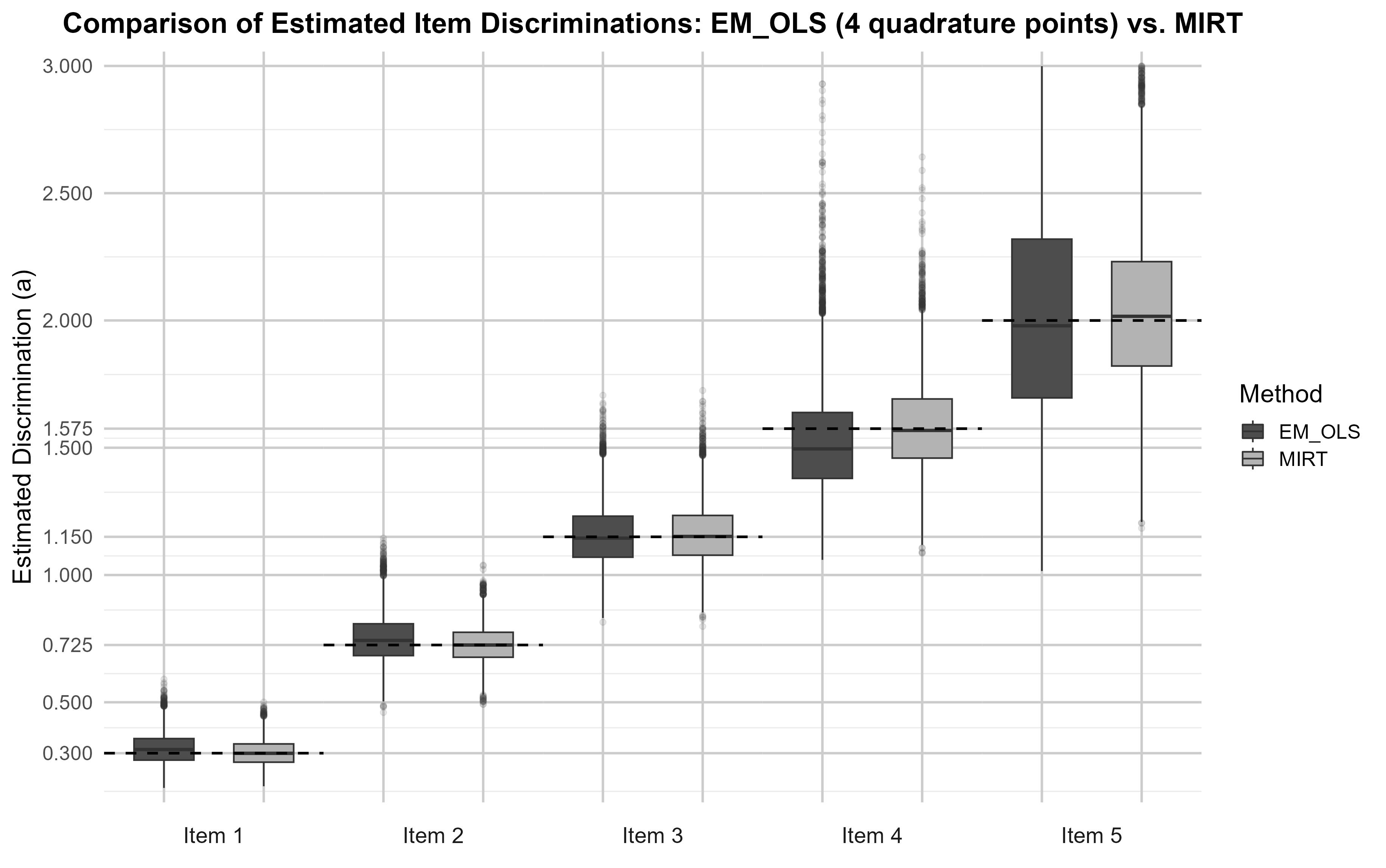}
\caption{Box plots of the estimates of the discrimination parameters for the \code{EM\_OLS}  and \code{mirt} procedures in the 2PL model}
\label{fig:OLSmirt2PLdiscr}
\end{figure}

\begin{figure}[h!]
\centering
\includegraphics[scale=.10]{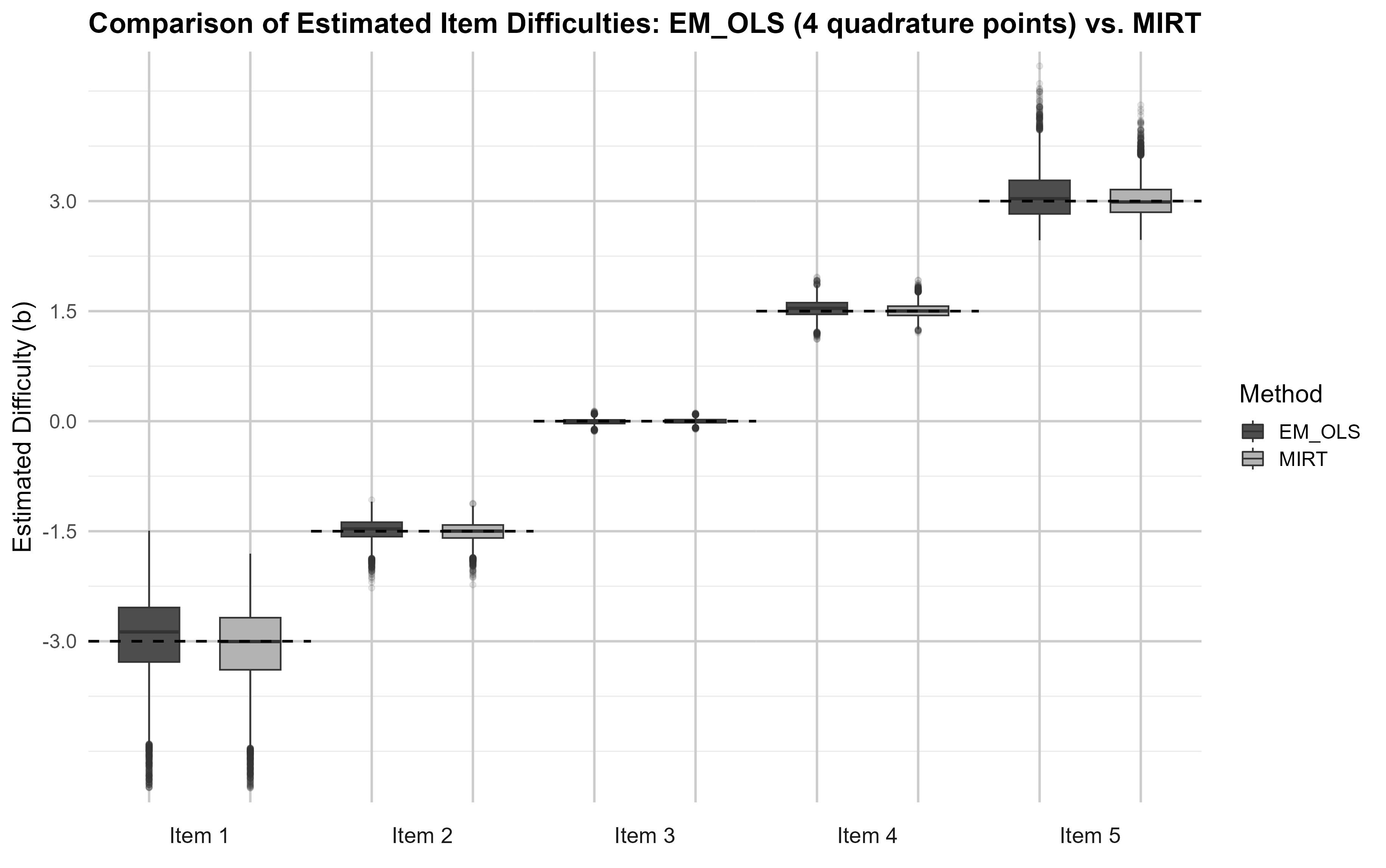}
\caption{Box plots of the estimates of the difficulty parameters for the \code{EM\_OLS}  and \code{mirt} procedures in the 2PL model}
\label{fig:OLSmirt2PLdiffs}
\end{figure}

Difficulty and discrimination parameters for data generation were respectively set to $b=\{-3,-1.5,0,1.5,3\}$ and $a = \{.3, .725, 1.15, 1.575, 2\}$ in order to cover that traditional ranges of item parameter magnitudes. Effects on the parameter estimates of the number of quadrature points were investigated for the following values $\{2,3,4,5,8,10,15\}$, results for this latter are reported in Appendix \ref{sec:quadsim}. Starting values were set to zero for all the difficulty parameters and to one for all discrimination parameters. It is therefore important to stress that both starting values and issues of practical identifiability as described in Section \ref{sec:completeIRT} might contribute to uncertainty in the parameter estimates of more extreme parameter magnitudes alongside the choice of quadrature points as described in Appendix \ref{sec:quadsim}. Given these parameters, 10,000 datasets were generated and estimated. Datasets were generated for samples of 5000 individuals in order to avoid estimation issues associated to small sample size. Figures \ref{fig:OLSmirt1PL}, \ref{fig:OLSmirt2PLdiscr}, and \ref{fig:OLSmirt2PLdiffs} compare the \code{EM\_OLS} and the \code{mirt} estimates for the 1PL and the 2PL models. The mean of the estimates alongside the Root Mean Squared Error (RMSE) and the number of cases in which outlier estimates were computed are given for all parameters in Tables \ref{tab:1PLest} and \ref{tab:2PLest}. Acceptable ranges outside of which parameter estimates were considered outliers are  $|b| < 5$ for the difficulty parameters and $.1<a<3$ for the discrimination parameters. It should be stressed that the results reported in the Figures have been filtered out for these values, while for completeness the Tables report unfiltered values of the mean estimates and of the RMSE.

As to the results, Table \ref{tab:1PLest} and Figure \ref{fig:OLSmirt1PL} report the comparison of the \code{EM\_OLS} procedure with \code{mirt} for the estimates of the difficulty parameters in the 1PL model. 

\begin{table}[h!]
    \centering
    \caption{Mean values and RMSEs of the difficulty parameter estimates for the \code{EM\_OLS} and \code{mirt} procedures in the 1PL model}
    \begin{tabular}{c c c c c c}\hline
        & Item 1 & Item 2 & Item 3 & Item 4 & Item 5  \\\hline
    True $b$  & -3.00 & -1.50 & .00 & 1.50 & 3.00\\
    \code{EM\_OLS} mean & -2.99 & -1.52 & .00 & 1.52 &  2.99 \\
     \code{EM\_OLS} RMSE  & .07 & .04 & .03 & .04 & .07\\
       \code{mirt} mean & -3.00 & -1.50 & .00 & 1.50 & 3.00 \\
    \code{mirt} RMSE & .07 & .04 & .03 & .04 & .06\\\hline
    \end{tabular}
    \label{tab:1PLest}
\end{table}

Table \ref{tab:1PLest} reports the mean value of the estimated difficulty parameters and the associated RMSE for both routines. Figure \ref{fig:OLSmirt1PL} reports the box-plots of the estimates of the difficulty parameters. For both routines in the 1PL case there were no outliers in the parameter estimates. As it can be noticed, the two routines are perfectly comparable.

Figures \ref{fig:OLSmirt2PLdiscr} and \ref{fig:OLSmirt2PLdiffs}, as well as Table \ref{tab:2PLest}, report instead  the comparison of the \code{EM\_OLS} procedure with \code{mirt} for the estimates of the discrimination and difficulty parameters in the 2PL model. Table \ref{tab:2PLest} reports the mean value of the estimated difficulty parameters and the associated RMSE for both routines. Figures \ref{fig:OLSmirt2PLdiscr} and \ref{fig:OLSmirt2PLdiffs} report the box-plots of the estimates of the discrimination and difficulty parameters. Box-plots have been given once the outlier estimates were filtered out while the Table reports the unfiltered results.

\begin{table}
    \centering
    \caption{Mean values, RMSEs, and number of outliers, for the the discrimination and difficulty parameter estimates for the \code{EM\_OLS} and \code{mirt} procedure in the the 2PL model}
    \begin{tabular}{c c c c c c}\hline
        & Item 1 & Item 2 & Item 3 & Item 4 & Item 5  \\\hline
    True $a$  & .300 & .725  & 1.150 & 1.575 & 2.000 \\
       \code{EM\_OLS} mean  &  .315 & .750 & 1.156 & 1.531 & 2.112 \\
     \code{EM\_OLS} RMSE  & .065 & .096 & .122 & .225 & .647\\
     \code{EM\_OLS} outliers  & 3 & 0 & 0 & 9 & 597 \\
    \code{mirt} mean  & .300 & .727 & 1.160 & 1.585 & 2.052\\
    \code{mirt} RMSE  & .054 & .072 & .119 & .179 & .335\\
    \code{mirt} outliers  & 2 & 0 & 0 & 0 & 112 \\\hline
     True $b$  & -3.00 & -1.50 & .00 & 1.50 & 3.00\\
      \code{EM\_OLS} mean & -2.98 & -1.48 & -.01 & 1.53 & 3.05\\
    \code{EM\_OLS} RMSE & .64 & .15 & .04 & .12 & .35 \\
    \code{EM\_OLS} outliers  & 91 & 0 & 0 & 0 & 0 \\
    \code{mirt} mean & -3.10 & -1.51 & .00 & 1.51 & 3.01 \\
    \code{mirt} RMSE & .61 &  .13 & .03 & .09 &  .24 \\
    \code{mirt} outliers  & 99 & 0 & 0 & 0 & 0 \\\hline
    \end{tabular}
    \label{tab:2PLest}
\end{table}

 Although the results show more instability of the \code{EM\_OLS} procedure w.r.t.\ \code{mirt} compared to the 1PL case, it appears that the performance of the OLS procedure is still comparable with the standard results provided by the \pkg{mirt} package.  Notice that, as it has been previously mentioned, both starting values and issues of practical identifiability as described in Section \ref{sec:completeIRT} might contribute to uncertainty in the parameter estimates of more extreme parameter magnitudes. As to the discrimination parameters it appears that the \code{EM\_OLS} procedure is more unstable in identifying higher magnitudes of the parameter. The number of outliers produced by the \code{EM\_OLS} procedure in the most extreme value ($a=2$) is five times greater than the number of outliers produced by the \code{mirt} procedure, but still small, being about the .06\% of the total cases. As to the difficulty parameters it appears that the two procedures are mostly comparable and both reveal more instability with the lowest magnitude of the parameter.

\section{Discussion}

The present contribution had a two-fold aim. On the one hand, to revise the fundamental notions associated to the item parameter estimation in 2 parameter IRT models from the perspective of the complete-data likelihood. Indeed, with the exception of notable works like \citet{WH1996}, most results for marginal maximum likelihood estimation of item parameters, following the wake of the seminal work of \citet{BA1981}, are given instead for the observed-data log-likelihood. On the other hand, we argued that, within an Expectation-Maximization approach, a sequence of closed-form estimators for discrimination and difficulty parameters can be obtained that simply corresponds to the Ordinary Least Square solution. As a proof-of-concept, the EM-based OLS procedure was implemented in the \proglang{R} environment. Simulations were used to explore the effects of the number of Gauss-Hermite quadrature points on the procedure and are reported in Appendix \ref{sec:quadsim}. As a result, it appears that two quadrature points provide reliable estimates for the 1PL while about 3 and 4 quadrature points provide reliable estimates for the 2PL. Parameter estimates returned by the EM-based OLS routine appear to be perfectly comparable with those returned by the package \pkg{mirt} \citep{C2012}.

These results have several possible consequences. Firstly, the EM-based OLS approach seems to lie in the intersection between the optimization methods traditionally used in IRT and those used in SEMs approaches to categorical variables. Indeed, the gist of the approach is to use the EM iterations to improve the estimates of the latent response variables that are then used to regress the IRT parameters. Secondly, gradient search methods might not be truly needed to estimate said parameters. Thirdly, albeit the EM procedure is well-known to be slow, the use of closed-forms can speed up the computational process and allow for potentially more scalability on larger set of items.

As to the limitations, although the results provided by the proposed EM-based OLS method are perfectly compatible with those provided by traditional gradient search methods, it appears that the OLS approach might be slightly more sensitive than traditional gradient search approaches to structural influences like the choice of starting parameters, the choice of the number (and method) of quadratures, and issues of practical identifiability. These factors might indeed lead to slightly more unstable estimates which manifest themselves as slight biases increasing with the actual magnitude of the parameters, higher RMSEs and variances of the estimators, as well as a higher numbers of convergences to outlier magnitude values of the parameters. This offers an interesting opportunity for future research to explore in more details the effects of these factors on the closed-forms estimators, thus also providing further insight on the reasons behind the instability of traditional gradient search methods solutions. Two directions are of particular interest. On the one hand, since the number of quadrature points in the Gauss-Hermite case dictates the range of the quadrature points, alternative methods like Gauss-Legendre quadrature might be considered to directly control the range of the latent trait values, while allowing to increase the number of quadrature points in such an interval, thus offering a possible way to further stabilize the estimates. On the other hand, these closed-form estimators appear by design to be sensitive to practical identifiability issues as they might converge to alternative set of values of the parameters that generate frequencies of responses that are very close to (and thus indistinguishable by given a certain amount of noise in the data) those generated by the true parameters. Since convergence to such a value would not correspond to convergence to a presumed local minima, but would actually correspond to convergence to the global minima of a different set of parameter values that simply generate a slightly different pattern of response frequencies, this offer the possibility to investigate practical identifiability issues that, up to our knowledge, have not been investigated so far in the IRT literature.

\section*{Acknowledgments}
This research was supported by the Deutsche Forschungsgemeinschaft (DFG) Grant No. NO 1505/2-1.

\appendix

\section{A note on the traditional IRT approach}\label{sec:marginalization}

The approach discussed by \citet{BA1981} and traditionally applied in IRT does not actually follow the complete-data log-likelihood but directly computes the partial derivatives of the observed-data log-likelihood \eqref{IRTlogdiscrete} in which the ability parameter is marginalized via Gauss-Hermite numerical quadratures. Partial derivatives w.r.t.\ the parameter $\gamma_i$ (which spans $a_i$ and $b_i$) for the item $i=j$ are then given by
\begin{align}
\frac{\partial \ell(\Gamma|\boldsymbol{X})}{\partial \gamma_j} & = \sum_{X\in \{0,1\}^{I}}{N_X}\frac{\partial}{\partial \gamma_j}\log{L(\Gamma|X)} = \sum_{X\in \{0,1\}^{I}}{N_X}\frac{1}{L(\Gamma|X)}\frac{\partial L(\Gamma|X)}{\partial \gamma_j}\nonumber\\
& = \sum_{X\in \{0,1\}^{I}}\frac{N_X}{\sum_{t=1}^T P(X|\theta_t, \Gamma)A(\theta_t; \mu, \sigma)}\sum_{t=1}^T\frac{\partial P(X|\theta_t, \Gamma)}{\partial \gamma_j}A(\theta_t; \mu, \sigma)\nonumber\\
& = \sum_{t=1}^T\left[\frac{\sum_{X\in \{0,1\}^{I}}N_X\frac{\partial P(X|\theta_t, \Gamma)}{\partial \gamma_j}}{\sum_{t=1}^T P(X|\theta_t, \Gamma)A(\theta_t; \mu, \sigma)}\right]A(\theta_t; \mu, \sigma).
\end{align}

As it holds for LI \eqref{local} that
the partial derivatives of $P(X|\theta_t, \Gamma)$ are
\begin{align*}
\frac{\partial P(X|\theta_t, \Gamma)}{\partial \gamma_j} & = \begin{cases}
 \frac{P(X|\theta_t, \Gamma)}{P_j(\theta_t, \Gamma_j)}\frac{\partial P_j(\theta_t, \Gamma_j)}{\partial \gamma_j} & X_j=1 \\
- \frac{P(X|\theta_t, \Gamma)}{1-P_j(\theta_t, \Gamma_j)}\frac{\partial P_j(\theta_t, \Gamma_j)}{\partial \gamma_j} & X_j=0 \end{cases}\\
& = \frac{(X_j-P_j(\theta_t, \Gamma_j))P(X|\theta_t, \Gamma)}{P_j(\theta_t, \Gamma_j)(1-P_j(\theta_t, \Gamma_j))}\frac{\partial P_j(\theta_t, \Gamma_j)}{\partial \gamma_j}
\end{align*}
than one has by substitution that
{\small\begin{align}\label{mirt}
\frac{\partial \ell(\Gamma|\boldsymbol{X})}{\partial \gamma_j} & = \sum_{t=1}^T\left[ \frac{N^1_{jt}-N_t P_j(\theta_t, \Gamma_j)}{P_j(\theta_t, \Gamma_j)(1-P_j(\theta_t, \Gamma_j))}\frac{\partial P_j(\theta_t, \Gamma_j)}{\partial \gamma_j}\right]
\end{align}}
where the quantities $N_{jt}, N_t$ are defined as
\begin{align} \label{patternbar}
N^{1}_{jt} = \sum_{X\in \{0,1\}^{I}}N_XX_jP(\theta_t|X,\Gamma) & \quad\text{and}\quad N_t = \sum_{X\in \{0,1\}^{I}}N_XP(\theta_t|X,\Gamma)
\end{align}
and where the posterior is defined by
\begin{align}\label{posteriortraditional}
  P(\theta_t|X,\Gamma) & =  \frac{P(X|\theta_t, \Gamma)A(\theta_t; \mu, \sigma)}{\sum_{t'=1}^T P(X|\theta_{t'}, \Gamma)A(\theta_{t'}; \mu, \sigma)}.
\end{align}

It is rather straightforward to notice that Equation \eqref{mirt} is exactly Equation \eqref{Qonettraditional}, and that the proportions \eqref{patternbar} as well the posterior \eqref{posteriortraditional} are exactly the same quantities defined in Equations \eqref{patternbargen} and \eqref{totalbargen} and by the posterior \eqref{posteriorIRT} but without the generalization to the $\nu_t$ parameters and the $n$-th step of the EM. Traditionally, the E-step consists in finding \eqref{patternbar} by treating $\Gamma$ as known while the M-step searches for the roots of \eqref{mirt}. Calculation of the derivatives for the item parameters yield
\begin{align}
\frac{\partial P_j(\theta_t, \Gamma_j)}{\partial a_j} & = (\theta_t-b_j)P_j(\theta_t, \Gamma_j)(1-P_j(\theta_t, \Gamma_j))\label{deriva} \\
\frac{\partial P_j(\theta_t, \Gamma_j)}{\partial b_j} & = -a_jP_j(\theta_t, \Gamma_j)(1-P_j(\theta_t, \Gamma_j))\label{derivb}
\end{align}
so that one has the system of equations
\begin{align*}
  \frac{\partial \ell(\Gamma|\boldsymbol{X})}{\partial a_j} & = \sum_{t=1}^T(\theta_t-b_j)\left[N^1_{jt}-N_t P_j(\theta_t, \Gamma_j)\right] =0 \\
   \frac{\partial \ell(\Gamma|\boldsymbol{X})}{\partial b_j} & = -\sum_{t=1}^Ta_j\left[N^1_{jt}-N_t P_j(\theta_t, \Gamma_j)\right] =0
\end{align*}
which is typically solved by means of computational methods like Newton-Raphson in order to devise the values of the parameters in $\Gamma_j$. As a final note, it is worth mentioning that the results discussed in Section \ref{sec:completeIRT} suggesting that there is a closed-form alternative to gradient search methods can already be noticed by observing that if a regression form of the 2 parameter IRT model is used, the previous system of equations would be rewritten as
\begin{align*}
  \frac{\partial \ell(\Gamma|\boldsymbol{X})}{\partial a_j} & = \sum_{t=1}^T\theta_t\left[N^1_{jt}-N_t P_j(\theta_t, a_j, \tau_j)\right] = 0\\
   \frac{\partial \ell(\Gamma|\boldsymbol{X})}{\partial \tau_j} & = \sum_{t=1}^T \left[N^1_{jt}-N_t P_j(\theta_t, a_j, \tau_j)\right] = 0
\end{align*}
in which the two parameters do not appear outside of the IRF $P_j(\theta_t, \Gamma_j)$. In particular since $N^1_j = \sum_{t=1}^T N^1_{jt}$ the second equation can be rewritten as $ N^1_j = \sum_{t=1}^T N_t P_j(\theta_t, \Gamma_j)$, which states that the proportion of all individuals that solved item $j$ is the sum over all ability levels of the proportions of individuals that solved the item in a given ability level. Such an interpretation already hints to the fact that rather than the sum being equal to zero, what is actual zero at every term is the single term in-between round brackets, so that $P_j(\theta_t, a_j, \tau_j) = \frac{N^1_{jt}}{N_t }$ is indeed an estimate of the IRF.

\section{A note on the estimator for the ability classes}\label{sec:memnership}

Following the same derivation as discussed by \citet{WH1996}, one can estimate the memberships parameters $\nu$ using the $Q_2$ term given by Equation \eqref{finitemixtureQ2}, that is
\begin{align}
  Q_2(\nu|\Gamma^{(n)}) = \sum_{X\in\{0,1\}^I}\sum_{t=1}^TN_XP(\theta_t|X,\Gamma^{(n)})\log{\nu_t}
  & = \sum_{t=1}^T N^{(n)}_{t}\log{\nu_t}\nonumber
\end{align}

Derivation w.r.t.\ the $\nu_t$ parameters can be carried out by means of Lagrangian multipliers to account for the normalization condition $\sum_t\nu_t=1$, thus introducing a further parameter $\lambda$, that is
\begin{align}
    \frac{\partial}{\partial \nu_l}(\sum_{t=1}^T N^{(n)}_{t}\log{\nu_t}+\lambda(\sum_t\nu_t-1)) & = \frac{N^{(n)}_{l}}{\nu_l}+\lambda,
\end{align}
which once set to null yields $N^{(n)}_{l} = -\lambda\nu_l$ and since $\sum_{l} N^{(n)}_{l} = -\lambda$ one gets
\begin{align}
    \hat{\nu}_l & = \frac{N^{(n)}_{l}}{\sum_{t'} N^{(n)}_{t'}}
\end{align}
which provides an estimate for the $l$-th probability of the latent ability class.

\section{Quadrature points}\label{sec:quadsim}

In the following simulations, the effect of the number of quadrature points (argument \code{n\_quads} in the \code{EM\_OLS} procedure) on the estimation of the parameters of the 1PL and 2PL models was explored. Difficulty and discrimination parameters for data generation were set to $b=\{-3,-1.5,0,1.5,3\}$ and $a = \{.3, .725, 1.15, 1.575, 2\}$ as in Section \ref{sec:simulations}. Starting values were set to zero for all the difficulty parameters and to one for all discrimination parameters. As already mentioned in Section \ref{sec:simulations}, both starting values and issues of practical identifiability as described in Section \ref{sec:completeIRT} can contribute to the uncertainty in the parameter estimates of more extreme parameter magnitudes. Nonetheless some effects of the number of quadrature points also emerge. These effects appear to be due to the fact that increasing the number of quadrature points within a Gauss-Hermite quadrature approach extends the interval of the quadrature points beyond the range of difficulties used to generate the data. As these quadrature points can be much greater or smaller than the actual difficulties, they are associated to latent ability classes $\nu_{t}$ for which it is difficult to estimate the latent variables $N_{Xt}$ over which the different proportions \eqref{patternbargen} and \eqref{totalbargen}. As these are used to generate the latent response variables $y^{(n)}_{jt}$, this affects parameter estimation. 

\begin{figure}[h!]
\centering
\includegraphics[scale=.10]{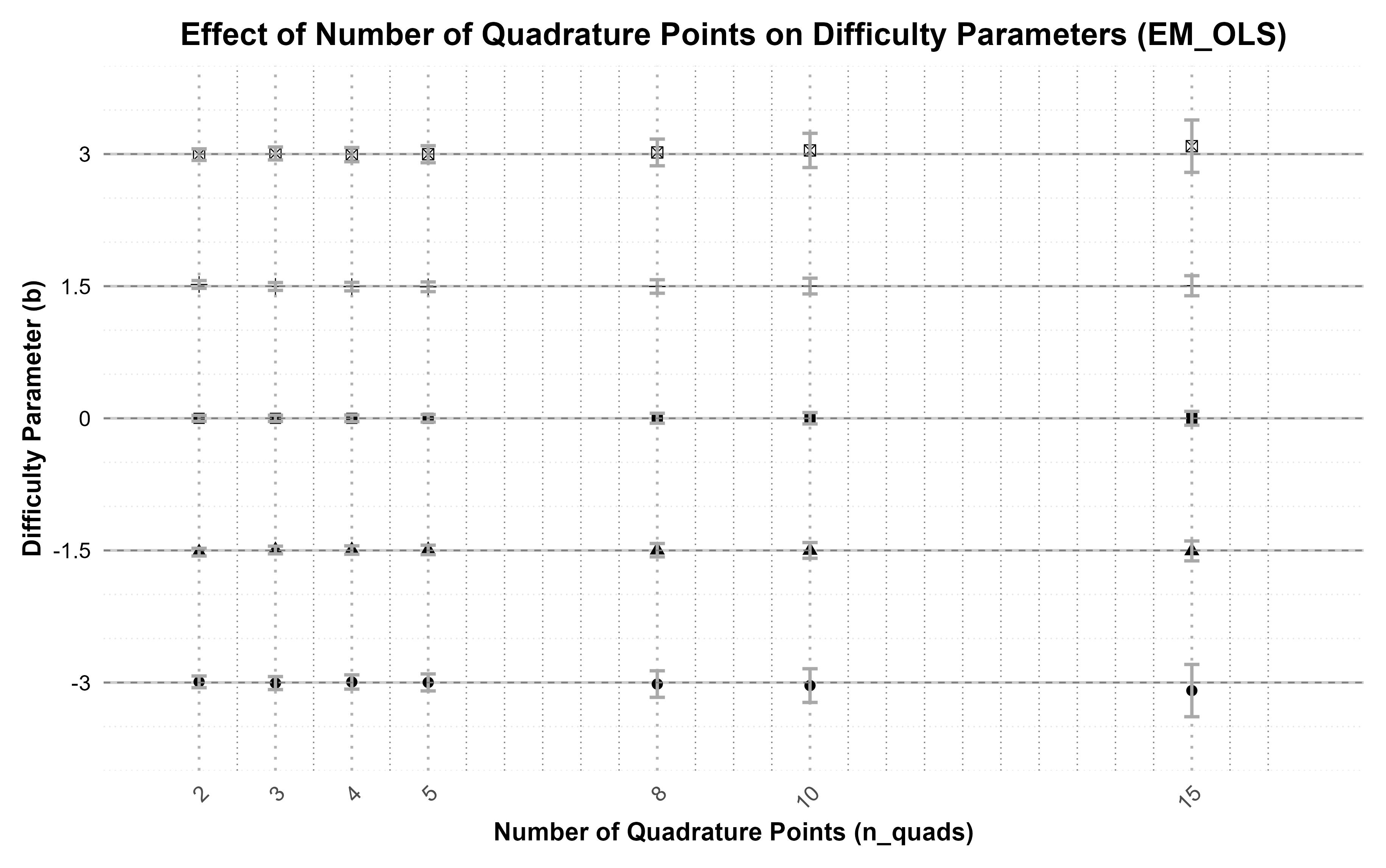}
\caption{Mean values and RMSEs of the difficulty parameters in the 1PL model for different choices of the number of quadrature points (\code{n\_quads})}
\label{fig:quads1PL}
\end{figure}

As it can be seen in Figure \ref{fig:quads1PL}, an increase in the number of quadrature points on the estimation of the 1PL model parameters implies both an increase in the uncertainty on the estimated parameters and the emergence of a small bias which is proportional to the magnitude of the estimated parameters. Two quadrature points appear to provide the best estimates for the difficulty parameter in the 1PL model.

\begin{figure}[h!]
\centering
\includegraphics[scale=.10]{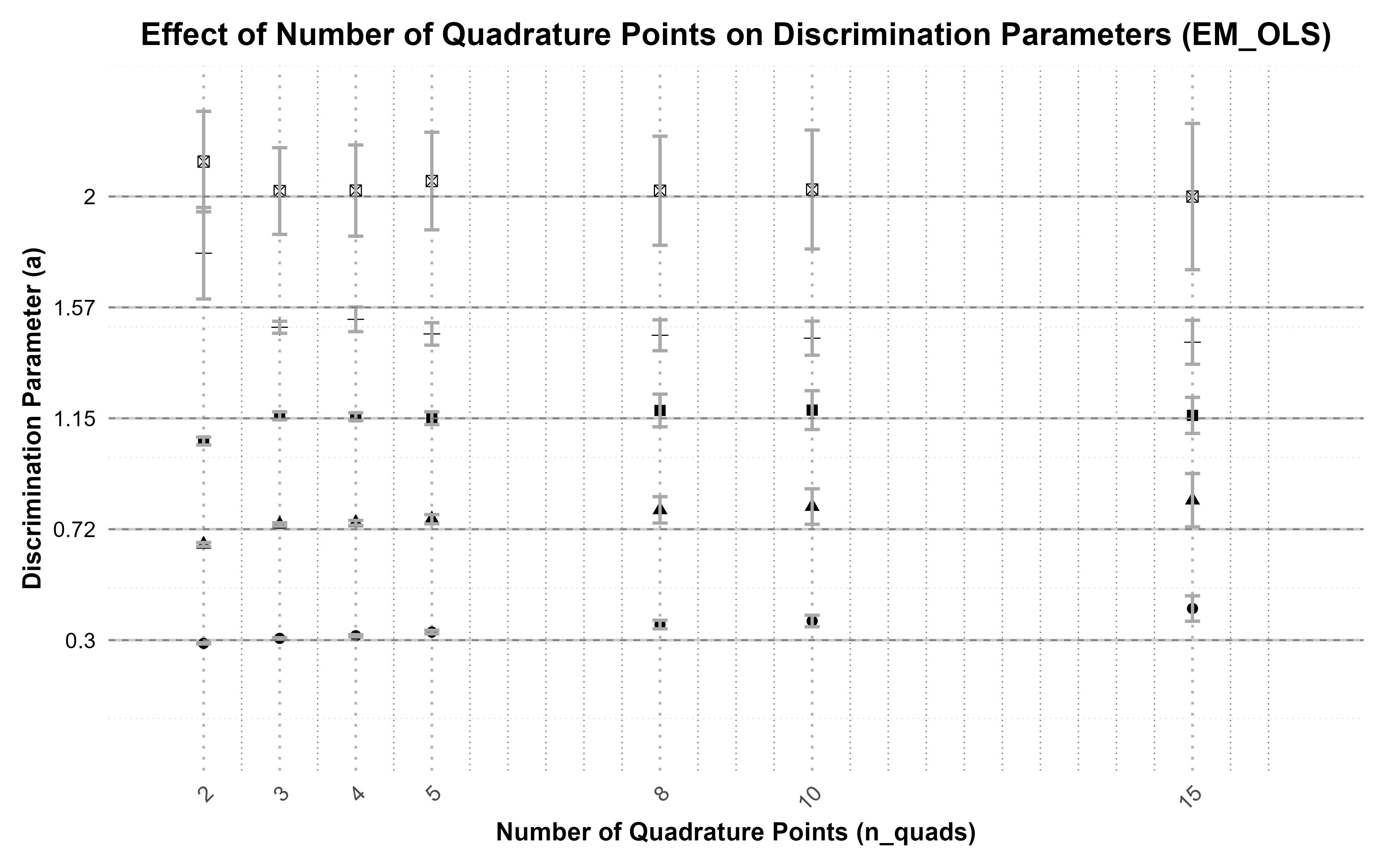}
\caption{Mean values and RMSEs of the discrimination parameters in the 2PL model for different choices of the number of quadrature points (\code{n\_quads})}
\label{fig:quads2PLdiscr}
\end{figure}

\begin{figure}[h!]
\centering
\includegraphics[scale=.10]{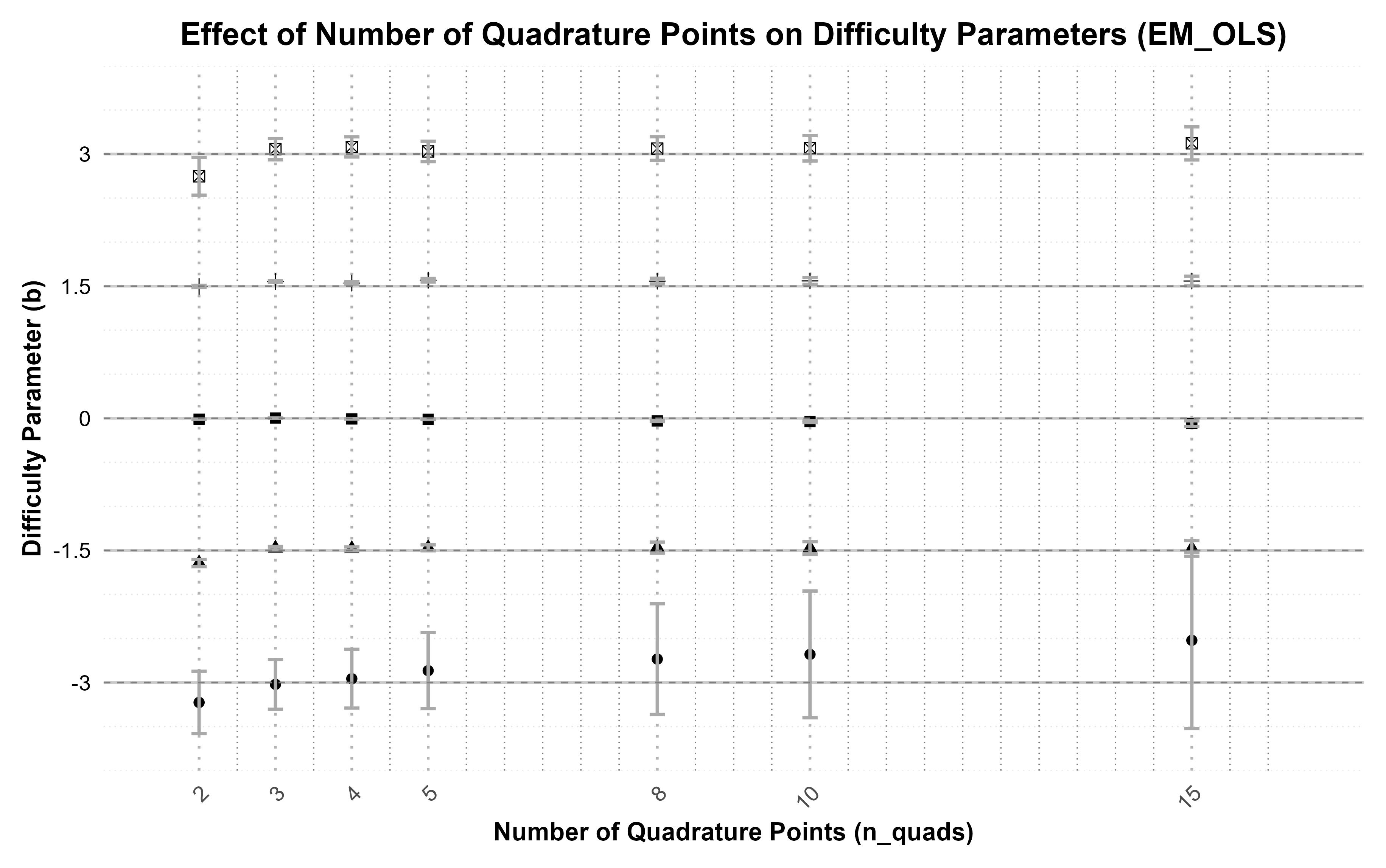}
\caption{Mean values and RMSEs of the difficulty parameters in the 2PL model for different choices of the number of quadrature points (\code{n\_quads})}
\label{fig:quads2PLdiffs}
\end{figure}

Similarly, as it can be seen in Figures \ref{fig:quads2PLdiscr} and \ref{fig:quads2PLdiffs}, an increase in the number of quadrature points on the estimation of the 2PL model discrimination and difficulty parameters also implies an increase in the uncertainty on the estimated parameters and the emergence of a small bias which is proportional to the magnitude of the estimated parameters. Interestingly the estimates at two quadrature points, which was the best scenario for the 1PL model, appear also to be biased. Three to four quadrature points appear to provide the best estimates for the difficulties and discrimination parameters in the 2PL. 

It is also evident from Figure \ref{fig:quads2PLdiscr} that there is an asymmetry between small and large magnitudes of the discrimination parameter. It appears indeed that a correct estimation of larger values of discrimination is more difficult to achieve independently on the number of quadrature points. Similarly, as it is evident from Figure \ref{fig:quads2PLdiffs}, in addition to the increase in bias and uncertainty for more extreme values of the difficulties, there is an asymmetry between large positive and negative magnitudes of the difficulties, with the latter that are more difficult to estimate.

\end{document}